\pdfoutput=1

  \documentclass[prl,two column,show keys,showpacs,superscriptaddress]{revtex4}

\usepackage{amssymb,amsmath}
 \usepackage{graphicx}
 \usepackage{epsfig}
 \usepackage{relsize} 

\newcommand{\gsim}
{\mbox{${~\raise.25em\hbox{$>$}\kern-.70em
\lower.25em\hbox{$\sim$}~}$}}
\newcommand{\lsim}
{\mbox{${~\raise.25em\hbox{$<$}\kern-.70em
\lower.25em\hbox{$\sim$}~}$}}
\newcommand{\beq}{\begin{equation}}
\newcommand{\beqa}{\begin{eqnarray}}
\newcommand{\eeq}{\end{equation}}
\newcommand{\eeqa}{\end{eqnarray}}

\newcommand{\eqn}[1]{eq.~(\ref{#1})}
\newcommand{\eqns}[2]{eqs.~(\ref{#1})-(\ref{#2})}

\begin{document}

\title{Spontaneous breaking of the flavor symmetry avoids the strong CP problem }

\author{Chee Sheng Fong}
    \email{chee.sheng.fong@lnf.infn.it}
 \affiliation{{\normalsize \it INFN, 
   Laboratori Nazionali di Frascati, 
   C.P. 13, I00044 Frascati, Italy}}

\author{Enrico Nardi}
  \email{enrico.nardi@lnf.infn.it}
 \affiliation{{\normalsize \it INFN, 
   Laboratori Nazionali di Frascati, 
   C.P. 13, I00044 Frascati, Italy}}
 \affiliation{{\normalsize \it 
 Instituto de F\'\i sica,
Universidad de Antioquia, A.A.{\it 1226}, Medell\'\i n, Colombia}}
\begin{abstract}
  \noindent
A promising approach to the Standard Model flavor puzzle 
is based on the idea that the $SU(3)^3$ quark-flavor
 symmetry  is spontaneously  broken by vacuum expectation values 
of `Yukawa fields'  which minimize   the  symmetry invariant scalar potential  
at  configurations  corresponding to the observed quark masses and 
mixing angles. We show that  this approach provides   a simple and elegant
explanation  for   $CP$ conservation in strong interactions.
\end{abstract}



\maketitle


A plethora of experimental verifications have confirmed QCD as the
correct theory of strong interactions. Still, QCD contains an
unresolved puzzle related to nonperturbative effects, which would
naturally lead to a neutron electric dipole moment orders of magnitude
larger than present limits. This is the strong-CP problem.  Basically,
the problem arises from the structure of the QCD vacuum~\cite{QCDvacuum}.
 Invariance under gauge transformations with nontrivial topological number requires 
that the physical  vacuum is a superposition of pure gauge configurations of 
all possible topological charges $p$, weighted by unimodular complex coefficients 
depending on   an angle $\theta$ whose value is not fixed by the theory: 
$|0\rangle_\theta = \sum_p e^{i\theta p}|p\rangle$.
Vacuum-to-vacuum transitions in the $\theta$ vacuum can be computed from the 
generating functional  by  adding to the naive Lagrangian a $\theta$-dependent  term:  
\beq
\label{eq:LQCD}
Z_\theta = \sum_{p=-\infty}^{+\infty} \int [d A_\mu]_p [d \phi ] [d \psi] [d\bar \psi] 
e^{i \mathlarger{\int} d^4 x \left\{ {\cal L}_{QCD} +{\cal L}_{\theta}\right\}}\,,
\eeq
where ${\cal L}_{\theta}=\theta \frac{g^2_s}{32\pi^2}  G^{a\mu\nu}\tilde G^a_{\mu\nu}$  with    $ G^{a\mu\nu}$ 
the gluon field strength tensor,   $\tilde G^a_{\mu\nu}$ its dual, and $g_s$ the strong coupling constant. 
Physics processes computed with   ${\cal L}_{QCD}$ alone are in striking agreement 
with experiments; however, the inclusion of  ${\cal L}_{\theta}$ clashes with 
phenomenology  because it  violates   $P$ and $CP$.   In particular, experimental bounds on CP-violation in 
strong interactions  are very tight, the strongest ones coming from the limits on the electric dipole moment of the neutron
$d_n < 0.29 \cdot 10^{-25}\,e$~\cite{Beringer:1900zz}  which implies $|\theta|< 0.56\cdot 10^{-10}$. 
Such a small number suggests that there should be a physical mechanism that enforces $\theta=0$. 

Under a chiral transformation  $\psi  \to e^{i\alpha\gamma_5} \psi$  the fermion path integral 
measure in \eqn{eq:LQCD}  is not invariant, but transforms  as~\cite{Fujikawa:1979ay}
\beq
\label{eq:dpsi}
\left[d\psi \right]
\left[\bar \psi \right] \to 
\left[d\psi \right]
\left[d\bar \psi \right]  e^{-i\mathlarger{\int} d^4x \frac{2n \alpha}{\theta}  {\cal L}_\theta}\,,
\eeq
where $n$ the number of quark flavors. 
For massless fermions ${\cal L}_{QCD}$ is invariant under a chiral transformation, 
which then  amounts to a  simple change of  variables  that cannot affect the physics. 
For $\alpha = \frac{\theta}{2 n} $ the  ${\cal L}_{\theta}$ term gets canceled,  implying that 
\eqn{eq:LQCD}   describes a theory equivalent to a theory with $\theta=0$.  
However, for the physical case of  massive quarks, a chiral transformation  modifies  the   
quark mass term  as  $\bar \psi\, M\, \psi \to \bar\psi \, M\,e^{2i\alpha\gamma_5} \,\psi$
with the net effect of shifting the overall phase of the determinant of the quark mass matrices:
\beq
{\rm Arg}\left[ \det \left(M_u M_d\right) \right]\to {\rm Arg} \left[\det \left(M_u M_d\right)\right] + 2 n\alpha.
\label{eq:argdet}
\eeq
 A non vanishing $\theta$ parameter can thus be brought back and forth 
from ${\cal L}_\theta$ to the fermions mass term; 
$\bar\theta \equiv \theta+ {\rm Arg} \left[\det \left(M_u M_d\right)\right])$   is however invariant under chiral
 transformations and represents a  physically meaningful parameter.   A CP conserving theory is obtained 
 when $\bar\theta=0$, and it  is equivalent to a theory 
in which $\theta=0$ and  the quark masses  are real.     


Another longstanding puzzle of the Standard Model (SM) is  related to  the peculiar way  
the quark flavor symmetry~\cite{MFV}  is broken: the observed breaking pattern involves unnaturally large
hierarchies between the quark masses, and the structure of the quark flavor mixing matrix, without any 
apparent explanation,  is very close to  diagonal. 
A promising  approach to explain this  puzzle is based on the idea that the SM Yukawa 
couplings are generated from dimension 5  operators involving quark bilinears,  the Higgs 
field,  and two  scalar  multiplets of `Yukawa fields', one for the up and one for the down sector:  
\begin{equation}
  \label{eq:Lm}
  -{\cal L}_m =  y_u\, \bar Q\, Y_u\, u\, \tilde H + 
 y_d\, \bar Q\, Y_d\, d\,  H \,, 
\end{equation}
where $Q$  is the  vector of the  left-handed quarks,  doublets of weak isospin,  $u,d$ denote 
  the triplets of  right-handed  $SU(2)$ quark singlets, $\tilde H = i\sigma_2 H^*$ with $H$ the Higgs field, 
  $Y_{u,d}$ are  $3\times 3$ matrices of complex scalar fields, and 
the couplings  $ y_{u,d}$  with dimension of an inverse mass are of order $1/\Lambda $,   
with   $\Lambda$  a large scale where the effective operators in
\eqn{eq:Lm} arise.  ${\cal L}_m$ is formally invariant under  transformations of the flavor  group ${\cal
  G}_{\cal F} = SU(3)_Q\times SU(3)_u\times SU(3)_d$   
  if we assign the fields to the   following representations of ${\cal  G}_{\cal F} $:
\begin{eqnarray}
  \label{eq:scalarsq}
  \nonumber
  Q\! \!&=&\!\! (3,\, 1\,,1),  \  \ \ u= (1,\, 3,\,1),  \  \  \  d= (1,\,1,\,3),  \\  
  H \! \!&=&\!\!  (1,\,1,\,1),  \  \  \ Y_u=(3\,,\bar 3,\,1),  \  \   Y_d = (3,\,1,\,\bar 3).  
\end{eqnarray}
The theoretical challenge is to find a ${\cal  G}_{\cal F} $-invariant scalar potential $V(Y_u,Y_d,Z)$ 
(where $Z$ denotes generically  additional scalars coupled to the Yukawa fields in
a symmetry invariant way) which can spontaneously break ${\cal  G}_{\cal F} $ yielding at the minimum 
of the potential a set of vacuum expectation values (vevs)  $\langle Y_{q}\rangle$ ($q=u,d$)   
with the observed structures of the SM Yukawa couplings.  
 In ref~\cite{Nardi:2011st}    it was found that the most general  renormalizable ${\cal  G}_{\cal F} $-invariant 
 scalar potential involving only $Y_u$ and  $Y_d$ admits the  tree-level vacuum configurations 
 $Y_{q} \sim {\rm diag}(0,0,v_{q})$. This appeared as a promising starting point to account for 
 the hierarchies $m_t\gg m_{c,u}$ and $m_b \gg m_{s,d}$ . However, in  ref.~\cite{Espinosa:2012uu} 
it was proven that the vanishing entries in $\langle Y_q\rangle $ cannot be lifted to non-vanishing values 
by any type of perturbative effects (loop corrections or higher dimensional operators involving 
 $Y_{u,d}$ only).   In the same paper it was also shown  that a hierarchical pattern 
 of Yukawa couplings with non-vanishing entries can  be obtained  by  including  
  additional ${\cal G}_{\cal F}$ multiplets.  As we will discuss in a forthcoming publication~\cite{inpreparation}, 
  the full structure of quark mixings and hierarchical masses   can in fact be obtained from  a 
  ${\cal  G}_{\cal F} $-invariant scalar potential   $V(Y_u,Y_d,Z)$ involving a suitable number of additional 
  scalars  $Z$. This  result proves that  spontaneous flavor symmetry breaking (SFSB)  is  a viable 
  approach to  tackle the SM flavor puzzle.
   In this paper  we show that  this same approach  avoids  the strong CP problem, since  
  the conditions $\theta = 0$ and $\det(M_u M_d) \in \mathbb{R}$ get  automatically  enforced in a 
  dynamical way.  It is quite  remarkable that this is an intrinsic   property of the  SFSB scenario  
  which avoids the need of  postulating  additional mechanisms to explain why  CP is conserved in strong interactions.

  The   Lagrangian terms relevant for our discussion are ${\cal L}_m$ in \eqn{eq:Lm} and the  SFSB  potential  
\beqa
\label{eq:V}
 V &=& V_1(Y_q) + V_2(Y_q,Z)+V_D\,,
 \eeqa
where $Y_q$ stands for ($Y_u,Y_d$). For simplicity 
we take at first all parameters to be real (see below for  generalization). 
$  {\cal L}_m$ in \eqn{eq:Lm}  is formally invariant under the (chiral) rephasing 
\beqa
\label{eq:qphase}
q_R &\to& e^{i\alpha} \ \  q_R\,, \qquad\qquad (q=u,d)\,, \\
\label{eq:Qphase}
Q_L &\to& e^{-i\alpha} \ Q_L\,, \\
\label{eq:Yphase}
Y_q &\to& e^{-2i\alpha}\  Y_q \,.
\eeqa
 $V_1$  contains all the ${\cal G}_{\cal F}$-invariant  renormalizable terms  involving 
 $Y_{q}$  with the exception of   terms involving the determinants of $ Y_q$, 
which are included in $V_D$.  These are all Hermitian monomials like 
${\rm Tr}( Y_qY_q^\dagger)$  and   ${\rm Tr}( Y_qY_q^\dagger Y_{q'}Y_{q'}^\dagger)$ 
($q,\,q'=u,d$) ~\cite{Nardi:2011st}  which  are trivially invariant  under  $Y_q$ rephasing. 
$V_2$ contains, besides  the Yukawa fields, other scalars $Z$ which are 
needed to  generate   through the   vevs $\langle Y_u\rangle$,  $\langle Y_d\rangle$ a set of 
 Yukawa couplings  in agreement with observations~\cite{Espinosa:2012uu,inpreparation}, 
and in general it  includes also non-Hermitian monomials.  However,  
by imposing suitable transformation properties for the $Z$ fields, one can ensure that also $V_2$ 
is formally invariant under  the transformation~\eqn{eq:Yphase} (see below). 

The ${\cal G}_{\cal F}$ invariant determinants
${\cal D}_q \equiv \det Y_q= \frac{1}{3!} \epsilon_{ijk}\epsilon_{lmn}(Y_q)_{il}(Y_q)_{jm}(Y_q)_{kn}$ 
 appear in $V_D$   that  reads: 
\beq
\label{eq:VD}
V_D=\sum_{q=u,d}  \left[\mu_q {\cal D}_q(x)+ {\rm h.c.}\right]   = \sum_{q=u,d} 2 \mu_q D_q(x)  \, \cos\left[\delta_q(x)\right],
\eeq
where $D_q=|{\cal D}_q|$  while $\delta_q(x)$ are the dynamical phases of the determinants 
of the matrices of Yukawa fields.  This is the only non invariant piece of the scalar potential, and  
under $Y_q$ rephasing transforms according to:
\beq 
\label{eq:delta}
\delta_q(x) \ \to \ \bar\delta_q(x) = \delta_q(x) -   n \alpha\,, 
\eeq
where   the  $Y_q$'s have  dimension   $n/2$, with $n$ the number of flavors. 
Given that  the  background values of the determinant phases $\langle \delta_q\rangle$  are 
modified  by the transformations \eqn{eq:Yphase} while all 
the other terms of the (classical) Lagrangian remain unaffected, 
we can  conclude that by themselves   ${\cal L}_m, 
V_1$ and $V_2$ leave the values of $\langle \delta_{u,d} \rangle$ undetermined. Therefore   only 
$V_D$ is responsible for fixing  their specific values at the potential minimum. 
By choosing $\alpha =  \frac{\theta}{2n}$, the term  ${\cal L}_{\theta}$ in \eqn{eq:LQCD} is canceled by the 
Jacobian of the fermion functional measure \eqn{eq:dpsi}, while minimization of $V_D$, 
that is obtained for $\cos\left[\bar\delta_q(x)\right] =-1$,  yields
\beq
\label{eq:deltaud}
\langle \bar\delta_q \rangle  =  \pi \ \    \Longrightarrow \ \  {\rm Arg}\left[ \det \left(\langle Y_u \rangle \langle Y_d\rangle \right) \right]= 0 \  ({\rm mod}\ 2\pi)\,.  
 \eeq
We see that the condition $\theta=0$ together with real quark masses  gets dynamically  realized  upon SSB of 
the flavor symmetry. This provides an elegant explanation for the non observation of strong CP violating 
effects~\footnote{Strictly speaking  ${\rm Arg}\left[ \det \left(\langle Y_q\rangle\right)\right]=\pi$  which means that 
both the  up and down determinants  are negative and we have   an even number  $2 n_{-}$  of negative mass 
eigenvalues . A pure chiral rotation (that is without transforming $Y_q$) can be used to change 
the sign of the negative masses. This will reinstate  a harmless  $\theta = 2 \pi n_{-}  = 0$  [ mod $2\pi$]. }.

Let us note at this point  that while the transformations \eqns{eq:qphase}{eq:Yphase}  are reminiscent of 
a $U(1)$ Peccei-Quinn (PQ) symmetry~\cite{Peccei:1977hh}, because of the non invariance of $V_D$   there is in fact no 
such a symmetry.   The PQ mechanism relies on a $U(1)_{PQ}$ Abelian symmetry that  is exact at the classical 
level, and gets  broken only by instanton effects  that generate  the analogous of our $V_D$ term. Spontaneous breaking 
of $U(1)_{PQ}$ implies that  a (quasi-)massless Nambu-Goldstone boson, the axion, must exist. 
In contrast, in the present case  there is no $U(1)_{PQ}$  and thus no axion.   Of course, if the flavor symmetry is global, a large  number of massless Nambu-Goldstone bosons will appear upon SFSB. They, however, can be sufficiently decoupled from  SM processes by taking the  SFSB scale sufficiently large~\cite{Wilczek:1982rv}. Alternatively, ${\cal G}_{\cal F}$    can be gauged, for example along the lines discussed in refs.~\cite{Albrecht:2010xh,Grinstein:2010ve,Guadagnoli:2011id}, and no new massless particles appear in the theory. 

In the remainder of the paper we will argue  that  the result \eqn{eq:deltaud}  (i)  is not affected by 
generalizing to complex Lagrangian parameters; 
(ii) it  does not get  spoiled by perturbative quantum corrections; (iii) it remains protected against 
possible  effects of  physics at scales $\bar \Lambda \gg \Lambda$, as for example Planck scale effects.

\medskip

(i) {\it \large Complex parameters.} \ \
Dropping the previous simplification of real parameters,  we generalize  \eqn{eq:Lm} 
 and  \eqn{eq:VD} allowing for $y_q \to y_q e^{i\vartheta_q}$  and $\mu_q \to \mu_q e^{i \varphi_q}$.
 After performing the chiral transformations \eqns{eq:qphase}{eq:Qphase}
we remove the phases from ${\cal L}_m$,  including also  $\vartheta_q$,   by  generalizing 
 \eqn{eq:Yphase} to: 
\beqa
\label{eq:Yphase2}
Y_q &\to & e^{-i(2\alpha +\vartheta_q)}\,  Y_q  \,. 
\eeqa
This   ensures that once we set $\alpha =\theta/(2n)$, if  $\langle \det Y_q\rangle \in \mathbb{R}$  
there is  no strong CP violation (complex phases in the Higgs vev can be always removed 
via $SU(2)\times U(1)$ rotations).

As we have said, $V_2(Y_q,Z)$ might contain some monomials which are not Hermitian.  
For example,  if we admit the set of fields $Z_Q,Z_u,Z_d$ transforming respectively 
in the fundamental representations of the simple subgroups 
$SU(3)_Q,\,SU(3)_u,\,SU(3)_d$ of ${\cal G}_{\cal F}$, and singlets 
under the remaining two,   we have~\cite{inpreparation}
 \beq
 \label{eq:V2}
 V_2 \ \supset\   \gamma_{ud}\, Z_{u}^{\dagger}Y_{u}^{\dagger}Y_{d}Z_{d} +\sum_{q=u,d} \nu_q Z_Q^\dagger Y_q Z_q +{\rm h.c.}\,, 
 \eeq
which is formally invariant under  (\ref{eq:Yphase2}) if we assign the following transformation rules to $Z_q$ and $Z_Q$:
 \beqa
\label{eq:Zqphase}
Z_q &\to&  e^{i\vartheta_q}\, Z_q\,,  \qquad \qquad\quad  (q=u,d)\,, \\
\label{eq:ZQphase}
 Z_Q &\to&  e^{-2i\alpha}\ Z_Q\,.
 \eeqa
 The additional complex phases  in $\gamma_{ud}$ and $\nu_q$ 
 appearing  in  \eqn{eq:V2}  will be  irrelevant in the following discussion.  
After a change of variables \eqn{eq:Yphase2} with $\alpha=\frac{\theta}{2n}$ , \eqn{eq:VD}  reads:  
\beqa
V_D &=& \sum_{q=u,d} 2\mu_q D_q (x) \cos\left[\bar\delta_d(x) \right]   \,, \\
 \bar\delta_q(x) &=& \delta_q(x) +\varphi_q  -  \frac{1}{2}(n  \vartheta_q +\theta)  \,. 
  \eeqa
Clearly the minimum of $V_D$ still occurs for $\langle \bar\delta_q\rangle =\pi$ 
and   ${\rm Arg}\left[\det(\langle Y_u\rangle \langle Y_d\rangle) \right]=
\langle \bar\delta_u\rangle+\langle \bar\delta_d\rangle = 0$  (mod $2\pi$) is preserved. 
\medskip

(ii) {\large \it Quantum corrections.}  The  one-loop Coleman-Weinberg corrections to the 
effective  potential~\cite{Coleman:1973jx,Jackiw:1974cv}  for a scalar field $Y=Y_q$ can be 
written,  in the $\overline{\rm MS}$ scheme, as 
\begin{eqnarray}
V^{loop} & = & \frac{1}{64\pi^{2}}\sum_{i}M_{i}^{4}\left[\log\frac{M_{i}^{2}}{\Lambda^{2}}-\frac{3}{2}\right],\label{eq:CW_1loop}
\end{eqnarray}
where the background dependent  mass functions $M_i^2$   are the eigenvalues of the matrix~\cite{Espinosa:2012uu} 
\begin{equation}
  \label{eq:M2ij}
[{\cal M}^2]_{ij,kl} = 
\frac{\partial^2 V(Y)}{\partial {\cal Y}_{ij} 
\partial {\cal Y}_{kl}}\,\bigg|_{\langle Y\rangle}\,, 
\end{equation}
where ${\cal Y}_{ij}=\{R_{ij},\,J_{ij}\}$ with $R_{ij}\,(J_{ij})=
\sqrt{2}\> {\rm Re}({\rm Im})Y_{ij} $, and the second derivatives are 
evaluated at the space-time independent background values 
$\langle Y\rangle$. 
 Here in $V(Y)$ we are including only the contributions to the
effective potential  coming from  self-interactions between
the components of the $Y$ field. This suffices for our scopes because these are the 
 only  terms  that carry a dependence on  the phase of  ${\cal D}$.  
The eigenvalues $M_i^2$ have been computed in \cite{Espinosa:2012uu} and
have the generic form: 
\beq
\label{eq:Mi2kl}
M_i^2  = C^2_i  +  {\cal F}_i \left(\mu {\cal D} + \mu^* {\cal D}^\dagger\right)\,,
\eeq
where $C^2_i$   are real coefficients 
which do not depend on the determinant phase 
$\delta(x)$,   and the real functions ${\cal F}_i$ sum up radical functions of  the argument which,
after the usual field redefinitions,  can be written as  $2\mu D \cos(\bar\delta)$. 
The derivative of the one loop potential 
\beq
\frac{\partial V^{loop}}{\partial \bar\delta } = 
\sum_i \frac{\partial V^{loop}} {\partial M^2_i}\> 
\frac{\partial M^2_i} {\partial \bar\delta } \propto \sin(\bar{\delta}) \,, 
\eeq
then sums up terms that all vanish at $\bar\delta =\pi$, which thus  extremizes all the 
$\bar\delta $ dependent one-loop contributions and ensures that the quark mass eigenvalues 
remain real.   Depending on the sign 
of   their specific  coefficients some of these
contribution will have a minimum in $\bar\delta =\pi$  while  others in $\bar\delta =0$. However, 
it is clear that the perturbative nature of these corrections cannot change the tree level global 
minimum   $\bar\delta =\pi$ into a local maximum,  so that  $\langle \delta_u\rangle +\langle \delta_d\rangle $ 
keeps vanishing  mod $2\pi$.

\medskip

(iii) {\large \it  Effective operators.} 
The PQ solution to the strong CP problem is not free from  some difficulties. 
The most serious ones are related to  Planck scale effects that, although suppressed by powers of the Planck mass, 
can induce sufficiently large $U(1)_{PQ}$ breaking effects to spoil the solution, unless the dimension 
of the operators is $\geq 10$~\cite{Kamionkowski:1992mf,Barr:1992qq}. 
It is then mandatory to study if a similar problem can occur also in the present case. 
Physics at a scale $\bar\Lambda \gg \Lambda$ might  generate  higher dimension effective operators of the form
\beq
{\cal O}^{(3n)} \sim \frac{1}{\bar\Lambda^{3n-4}} D^n \cos\left[\phi_{\bar\Lambda} + n \bar\delta\right], 
\eeq
with $n>1$, which, by   bringing  in   new complex phases $\phi_{\bar\Lambda}$ ,    could  shift the  minimum
away from $\bar\delta=\pi$.  Let us investigate  which are the necessary  conditions for this to occur. We assume that 
a new minimum develops at $\bar \delta = \pi -\epsilon$. At this minimum, the tree level potential 
is then lifted by  
\beq
\Delta V \equiv  \left[V(\pi-\epsilon)\right]-
\left[V(\pi)\right]=\mu  \epsilon^2\langle D\rangle \,. 
\eeq
The leading effective term induced by the new physics must be of the form:
\beq
V_{\bar\Lambda} = \frac{D^2}{\bar\Lambda^2} f(\phi_{\bar\Lambda},\bar\delta)\,,
\eeq
and we take conservatively    a change   in   $V_{\bar\Lambda}$   linear in $\epsilon$:
\beq
\Delta V_{\bar\Lambda}\equiv
V_{\bar\Lambda}(\pi-\epsilon)-
V_{\bar\Lambda}(\pi) = - \frac{\langle D\rangle^2}{\bar\Lambda^2}\epsilon.
\eeq
The condition for the minimum to move away from  $\bar\delta=\pi$ is that  
\beq
\Delta V + \Delta V_{\bar\Lambda} <0\,, 
\eeq
which yields:
\beq
\label{eq:epsilon}
\epsilon < \frac{\langle D\rangle}{\mu\bar\Lambda^2}\sim \> 10^{-6} \, \frac{\Lambda^2}{\bar\Lambda^2}\,,
\eeq
where we have used $\frac{\langle D\rangle }{\Lambda^2\mu}= \frac{m_u}{\Lambda}
\frac{m_c}{\Lambda} \frac{m_t}{\mu}\sim  10^{-6}$\,.
Conflicts with phenomenology would in fact  occur only if the  shift  away from the minimum in $\pi$ 
induces a too large neutron electric dipole moment, which requires 
 $\epsilon\gsim 10^{-10}$.   From this and from \eqn{eq:epsilon}  we obtain that 
 the new scale $\bar \Lambda$ cannot be too far from the scale of flavor symmetry breaking:  $\bar\Lambda  < 10^{2} \Lambda$ 
(since the value of the top quark Yukawa coupling requires $\langle {\rm Tr}(Y_u)\rangle / \Lambda\sim 1$ , 
we can take  the flavor breaking scale  to be of order $\Lambda$).
In particular,  when   $\bar\Lambda = M_{\rm Planck}$,   whenever  the  flavor breaking scale  
is not unreasonably large  (say, $\Lambda \lsim  10^{17}\,$GeV)   we should not worry about the possibility 
that Planck scale effects could reintroduce  dangerous strong CP violating effects.   

Symmetry invariance under the quark flavor group ${\cal G}_{\cal F}=SU(3)_Q\times SU(3)_u\times SU(3)_d$ 
requires promoting  the  quark Yukawa couplings $Y_q$ to dynamical fields transforming properly under ${\cal G}_{\cal F}$. 
The  minimum of the   ${\cal G}_{\cal F}$-invariant    potential should then  break  ${\cal G}_{\cal F}$ 
spontaneously, producing a symmetry breaking pattern   for $Y_q$  in agreement  with observations~\cite{inpreparation}. 
We have shown that the determinants  of the Yukawa matrices  $Y_{u,d}$ are the only terms in the scalar potential that  are 
not  invariant under a phase transformation of the Yukawa fields.   By a change of field variables, the QCD P and CP violating 
$\theta$ term can be moved to the quark-Yukawa Lagrangian, and from there to the Yukawa determinants. We have 
shown that at the minimum of the SFSB  potential  $\det(Y_q)$  acquires a real  vev,  
which yield real quark masses that can be made all positive without reinstating a  $\theta$ term. 
Thus, a general property of breaking spontaneously the quark-flavor symmetry is to automatically account 
for the non-observation of strong-CP violating phenomena.
Of course, since a  real and positive $\det(Y_q)$  is obtained for example from the  
parametrization $Y_q = V_q Y^{\rm diag}_q$ with $Y^{\rm diag}_q$ diagonal with real and positive 
entries, and $V$ a complex special unitary  matrix  ($\det V=+1$), CP violation in the electroweak 
sector can still occur. An example of a SFSB potential $V(Y_u,Y_d,Z)$ realizing   this scenario  will be presented 
in a forthcoming paper~\cite{inpreparation}.


\end{document}